\documentclass[11pt,a4paper]{article}

\usepackage[utf8]{inputenc}
\usepackage[T1]{fontenc}
\usepackage[english]{babel}
\usepackage{amsmath, amssymb}
\usepackage{graphicx}
\usepackage{dcolumn}
\usepackage{doi}
\usepackage{hyperref}
\usepackage{authblk}
\usepackage{geometry}
\usepackage{caption}
\usepackage{subcaption}
\usepackage{abstract}
\usepackage[numbers,sort&compress]{natbib}
\usepackage{color}

\begin{document}


\newcommand{\noyA}[2]{\rm{^{#2}#1}}
\newcommand{\elratio}[3]{R_{\rm el}(\noyA{#1}{#2}/\noyA{#1}{#3})}
\newcommand{\inelratio}[2]{R_{\rm inel.1}(\noyA{#1}{#2})}
\newcommand{\AOMP}{AOMP}
\newcommand{\aomp}[1]{$aop#1$}
\newcommand{\MSM}{$aop2$-M}
\newcommand{\MSBa}{$aop2$-Ba}
\newcommand{\MSBb}{$aop2$-Bb}
\newcommand{\MSBc}{$aop2$-Bc}
\newcommand{\MSMBa}{$aop2$-MBa}
\newcommand{\MSMBb}{$aop2$-MBb}
\newcommand{\MSMBc}{$aop2$-MBc}
\newcommand{\AtoM}{$aop9$-M}
\newcommand{\MSTNda}{$aop2$-T76-142}
\newcommand{\MSTNdb}{$aop2$-T76-146}
\newcommand{\WS}{WS}
\newcommand{\DF}{DF}
\newcommand{\rMS}{r_o}
\newcommand{\aMS}{a_o}
\newcommand{\Rag}{R_{(\alpha,\gamma)}}
\newcommand{\Ran}{R_{(\alpha,n)}}


\title{\textbf{Studies of $\noyA{Sm}{144,148}+\alpha$ potential for the $p$-process nucleosynthesis} 
}%

\author[1]{C.~Soto}
\author[1]{C.~Ducoin}
\author[1]{N.~Millard-Pinard}
\author[1,2]{B.~M.~Rebeiro}
\author[1]{O.~St\'ezowski}
\author[3]{N.~de~S\'er\'eville}
\author[3]{F.~Hammache}
\author[1]{A.~Chalil}
\author[1]{Y.~Demane}
\author[3]{C.~Bachelet}
\author[2]{J.-C.~Thomas}
\author[3]{M.~Assi\'e}
\author[3]{M.~Benhatchi}
\author[3]{V.~Girard-Alcindor}
\author[2]{S.~V.~Harissopulos}
\author[3]{H.~Jacob}
\author[4]{A.~Lagoyannis}
\author[3]{S.~Morard}
\author[5]{S.~Nandi}
\author[5]{J.~P\'epin}
\author[3]{L.~Perrot}
\author[6]{A.~M. S\'anchez-Ben\'{\i}tez}
\author[3]{I.~Stefan}
\author[3]{T.~Zanatta-Martinez}

\affil[1]{Université Lyon 1, CNRS, IP2I, UMR 5822, Villeurbanne, France}
\affil[2]{Grand Acc\'el\'erateur National d'Ions Lourds (GANIL), CEA/DRF-CNRS/IN2P3, Caen, France}
\affil[3]{Université Paris-Saclay, CNRS/IN2P3, IJCLab, 91405 Orsay, France}
\affil[4]{Tandem Accelerator Laboratory -- CALIBRA, Institute of Nuclear and Particle Physics, NCSR ``Demokritos'', Aghia Paraskevi-Athens, Greece}
\affil[5]{SUBATECH, IMT Atlantique, Universit\'e de Nantes, France}
\affil[6]{Dept. Ciencias Integradas, Facultad de Ciencias Experimentales, Centro de Estudios
Avanzados en F\'{\i}sica, Matemática y Computación, Unidad Asociada GIFMAN,
CSIC-UHU, Universidad de Huelva, 21071, Huelva, Spain.}

\maketitle


\begin{abstract}
Nucleosynthesis reaction networks leading to $p$-nuclei involve a combination of different types of photodisintegration and capture reactions, as well as $\beta^+$ decays or electron captures. Photodisintegration reactions involving $\alpha$ particles present a particular interest as they serve as branching points of the reaction networks. The cross sections of these reactions depend crucially on the $\alpha$-nucleus interaction. The $\alpha$ optical model potential ({\AOMP}) is determined  mostly by means of experimental differential elastic scattering distributions. Several previous studies have focused on the case of $\noyA{Sm}{144}$, an intriguing $p$-nucleus that is semi-magic with 82 neutrons. This work presents new experimental data on $\alpha$ elastic and inelastic scattering on $\noyA{Sm}{148}$, its closest stable isotope. Isotopic effects on the description of the {\AOMP} are studied, as well as their consequences on the prediction of $\alpha$-induced reaction cross sections at astrophysical energies.
It is shown that the isotopic ratio for $(\alpha,\gamma)$ cross sections can be multiplied up to a factor of two when these effects are included. 
\end{abstract}

\maketitle


\section{Introduction}

Reactions involving $\alpha$ particles play a key role in many nucleosynthesis processes, including large reaction networks with thousands of exotic nuclei occurring in explosive events. One of these situations is the $p$-process~\cite{Arnould-03,Rauscher-13}, producing around 30 stable nuclei between $\noyA{Se}{74}$ and $\noyA{Hg}{196}$, called $p$-nuclei: 
these nuclei are neutron deficient and have very low isotopic abundances. 
They are shielded from the dominant processes that produce nuclei beyond iron, namely the $s$- and $r$-processes, 
which alternate between neutron captures and $\beta$-decays~\cite{Kappeler-11,Arnould-07}.
The main scenario to explain the production of these $p$-nuclei is called 
$\gamma$-process~\cite{Rayet-95,Travaglio-15}. 
It starts from seed nuclei produced beyond iron by neutron captures, which then encounter very high temperatures (a few GK) during a supernova explosion. 
At such temperatures, photodisintegrations first remove neutrons by $(\gamma,n)$ reactions until reaching the neutron-deficient part of the nuclear chart, then the reactions proceed mostly with $(\gamma,p)$, $(\gamma,\alpha)$, $(n,\gamma)$, and in the low-mass region, $(p,\gamma)$~\cite{Rauscher-13}.

The modeling of such reaction network requires theoretical calculations based on the Hauser-Feshbach statistical formalism~\cite{Hauser-Feshbach}, giving cross sections that depend especially on the particle transmission probabilities, nuclear level densities and $\gamma$ strength functions. The $\alpha$ transmission is usually determined in the framework of the optical model, using an $\alpha$-nucleus optical model potential ({\AOMP}) made of a real part, which accounts for elastic scattering, and an imaginary part responsible for all non-elastic channels. 
Many efforts are devoted to the improvement of a global {\AOMP} that would be reliable for a large range of nuclei, at energies relevant to astrophysical processes. In the case of $(\alpha,\gamma)$ reactions in the $p$-process, the corresponding Gamow peak is typically around 5 to 10 MeV.

In a pioneering work by McFadden and Satchler in 1966~\cite{aomp2}, the authors analysed $\alpha$ scattering data at 24 MeV on 19 nuclei from oxygen to uranium to obtain the mass-dependent parametrization of a Woods-Saxon {\AOMP}. This work has been widely used in nucleosynthesis calculations. However, in 1997, a measurement of the $\noyA{Sm}{144}(\alpha,\gamma)\noyA{Gd}{148}$ reaction in the 10-13 MeV energy range by Somorjai \textit{et al.}~\cite{Somorjai-97} attracted attention on the inadequacy of existing {\AOMP} to reproduce the $\alpha$ reactions at astrophysical energies, predicting in this case far too large cross sections. Modern attemps to define a global {\AOMP} suitable for nuclear astrophysics are now made compatible with this constraint, either through an energy dependance of the parameters~\cite{aomp345,aomp6}, or by defining an imaginary part that allows to reproduce the results of a simpler barrier transmission model for the determination of the $(\alpha,\gamma)$ cross section~\cite{mohr_2020}. 

The case of $\noyA{Sm}{144}$, among the $p$-nuclei, has attracted a lot of attention even before the challenging results by Somorjai, since the $\noyA{Sm}{146}/\noyA{Sm}{144}$ production rate was involed in a method proposed in~\cite{Audouze-72} to characterize the chronometry of $p$-process contributions to the solar-system composition. In order to determine the $\alpha-\noyA{Sm}{144}$ {\AOMP}, in 1997, Mohr \textit{et al.}~\cite{Mohr-97} measured with high accuracy the $\noyA{Sm}{144}(\alpha,\alpha)$ elastic scattering cross section at 20 MeV. A more recent measurement was performed by Kiss \textit{et al.} in 2022~\cite{Kiss-22}, confirming these previous results and extending them down to 16.13 MeV. In such elastic-scattering measurements, the angular distribution at backward angles is the most impacted by the nuclear interaction, making this angular region particularly important to benchmark the models. Concerning the $(\alpha,\gamma)$ reactions, the results by Somorjai have been confirmed in 2020 by Scholtz \textit{et al.}~\cite{Scholtz-20}, using a similar activation technique, with an improved $\alpha$ detection system to measure the $\noyA{Gd}{148}$ activity.
Even more recently, the ${\noyA{Sm}{144}}(\alpha,n){\noyA{Gd}{147}}$ cross section was measured in 2023 by Gyürky \textit{et al.}~\cite{Gyurky-23} also with the purpose of improving the description of the {\AOMP} and its ability to predict the $(\alpha,\gamma)$ cross section at astrophysical energies. 

Given that $\noyA{Sm}{144}$ is a semi-magic nucleus ($N=82$),
we complement in this work the existing efforts focused on this nucleus with a study of its closest stable neighbour, $\noyA{Sm}{148}$, in order to examine the role of magicity on the {\AOMP} properties that are extracted. 
Isotopic effects in $\alpha$ elastic scattering 
were suggested to be a key observable to constraint the {\AOMP}, for instance by Palumbo \textit{et al.}~\cite{Palumbo-12} who studied the tellurium isotopic chain from $\noyA{Te}{120}$ to $\noyA{Te}{130}$. Concerning samarium, little attention has been given to the {\AOMP} of the different isotopes since the work performed by Badawy \textit{et al.} in 1978~\cite{Badawy-78}, in which the excitation function of the elastic $\alpha$ scattering close to  180$^\circ$ was analyzed for several nuclei, including $\noyA{Sm}{144,148,150,152}$.

In the present work, we performed a measurement of $\alpha$ elastic scattering angular distribution at 20 MeV (beam energy) on $\noyA{Sm}{144}$ and $\noyA{Sm}{148}$ in order to compare the properties of the {\AOMP} for the two isotopes, semi-magic and non-magic, and check the reproduction of existing data on $\noyA{Sm}{144}$. 
The experiment was performed at the ALTO facility (Accélérateur Linéaire et Tandem d'Orsay) 
of IJCLab (Laboratoire de Physique des 2 Infinis Irène Joliot-Curie),
using the Split-Pole magnetic spectrometer~\cite{SPENCER1967181}, 
with targets of high isotopic purity. 
We will show that relative observables comparing the two isotopes are of great interest for the characterization of the {\AOMP}. 

The experimental setup is described in Section~\ref{sec-setup}, 
target production and characterization in Section~\ref{sec-targets}.
The data analysis procedure is explained in Section~\ref{sec-analysis}, and Section~\ref{sec-discussion} is dedicated to the discussion, comparing the elastic and inelastic data we obtained with the predictions of different models calculated with the TALYS-2 reaction code~\cite{talys}. Standard {\AOMP} models available are used, and some {\AOMP} models adapted to the comparison of the two samarium isotopes are built. The impact of isotopic effect on $(\alpha,\gamma)$ cross sections at astrophysical energies, and on $(\alpha,n)$ cross sections below 20 MeV, is also discussed.


\section{\label{sec-setup}Experimental setup}

The experiment was performed at the ALTO facility, located at IJCLab in Orsay, France. The accelerator used for the present measurement is the 15 MV Tandem Van de Graaff, capable of delivering high-intensity and stable ion beams for nuclear physics experiments. In this study, a mono-energetic beam of $\alpha$ particles was accelerated to a laboratory energy of 20~MeV. The typical beam current ranged from 20~enA to 100~enA, 
optimized to ensure high statistics while minimizing damage on the targets and pile-up effects in the detectors.

\begin{figure}
    \centering
    \includegraphics[width=1.0\textwidth]{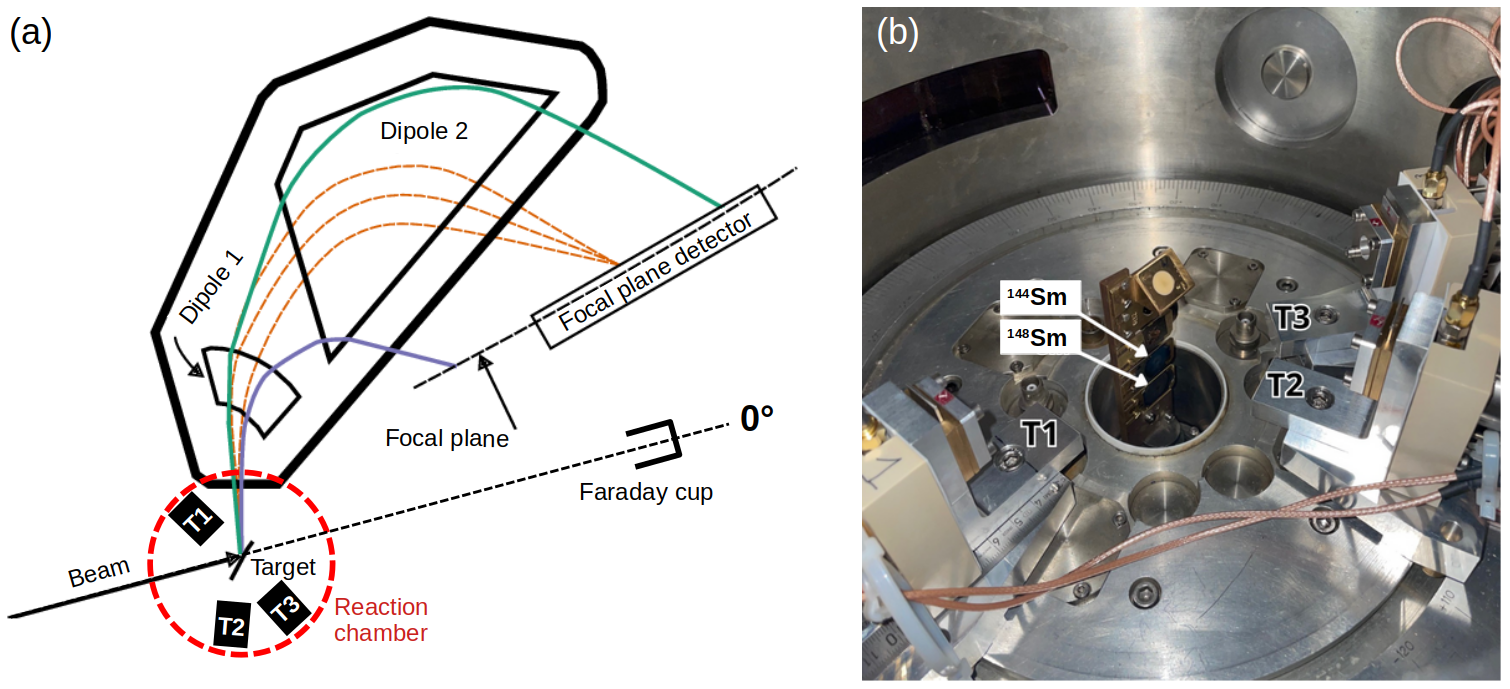}
    \caption{\label{fig:chambre} (a) Sketch of the experimental setup illustrating the Split-Pole spectrometer and the three silicon-detector telescopes $T_1$, $T_2$, and $T_3$ (adapted from \cite{Marshall2018TheFD}). (b) Picture of the scattering
chamber showing telescopes $T_1$, $T_2$, $T_3$ and the samarium targets in the center.}
\end{figure}

The $\alpha$ beam was directed onto an isotopically enriched samarium target, either $\noyA{Sm}{144}$ or $\noyA{Sm}{148}$, mounted at the center of a high-vacuum scattering chamber. To perform angular distribution measurements, we employed a hybrid detection system combining high-resolution magnetic analysis by the Split-Pole spectrometer and silicon-based charged-particle spectroscopy, as shown in Figure~\ref{fig:chambre}.

The Split-Pole spectrometer of ALTO 
uses three complementary detectors situated at its focal plane.
The first one is a position-sensitive gas detector, which gives the particle position along the focal plane and therefore its magnetic rigidity $B\rho$. The second one is a gas proportional counter measuring the energy loss $\Delta E$, used for particle identification. The last detector (not used in our analysis) is a plastic scintillator, placed at the end of the focal plane, which measures the residual energy of the particle. The spectrometer provides precise momentum analysis through magnetic rigidity selection, and detects the scattered particles with an excellent energy resolution ($E/ \Delta E  \approx 2000$). 
Due to mechanical constraints, the maximal angle we could reach with Split-Pole was $85^\circ$. 

\begin{figure}[htbp]
    \centering
    \includegraphics[width=1.0\textwidth]{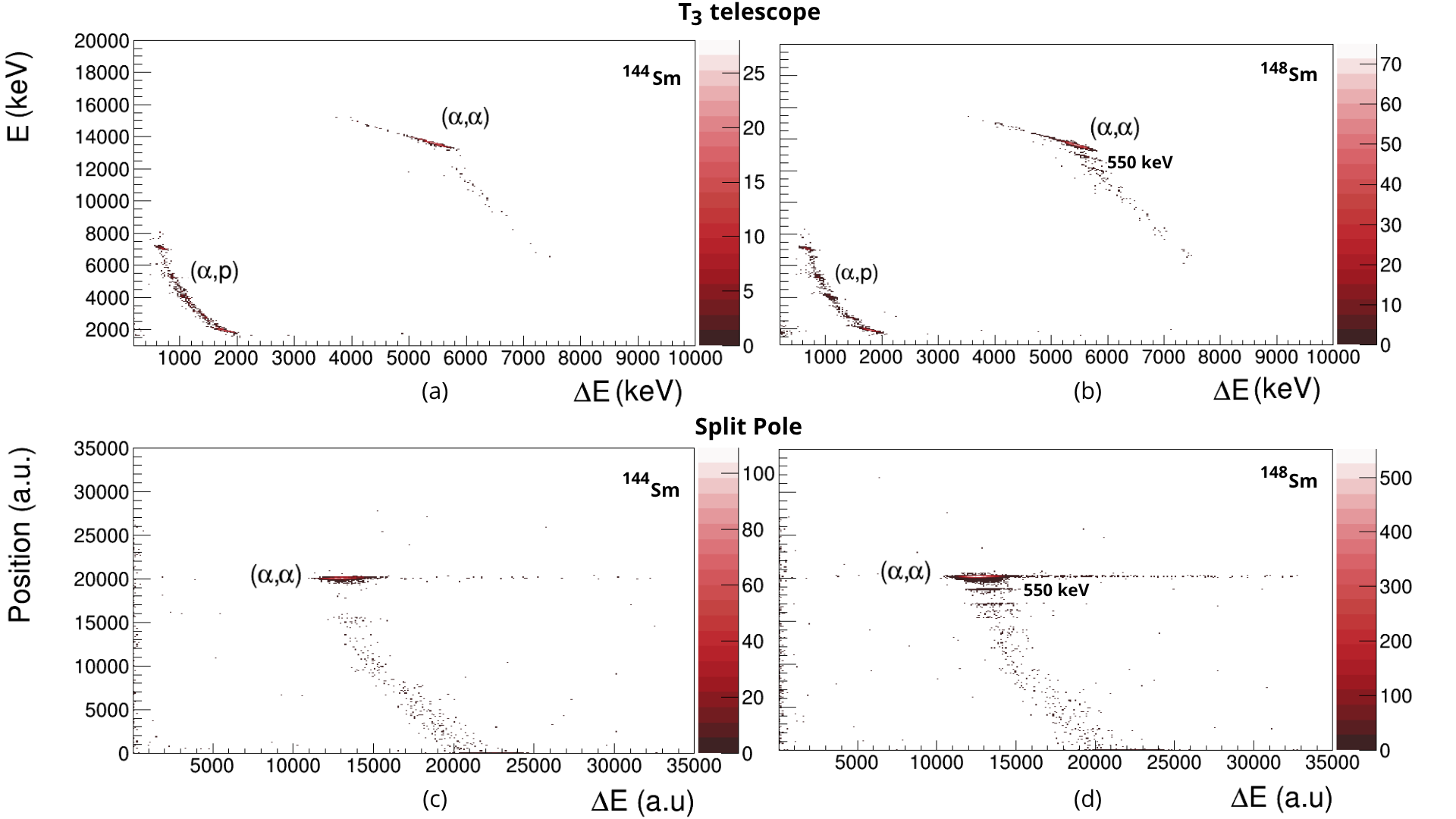}
    \caption{\label{fig:e-de_plot}
    Identification plots of the reaction channels measured 
    for a 20 MeV $\alpha$ beam incident on two samarium targets 
    ($\noyA{Sm}{144}$ on the left, $\noyA{Sm}{148}$ on the right).
    Upper panels: $E-\Delta E$ spectra measured with telescope $T_3$ placed at $109^\circ$ with respect to the incident beam direction. 
    On these plots, in addition to $\alpha$-scattering channels,
    $(\alpha,p)$ reaction channels are present, due to reactions on the backing material and target contaminants, as discussed in Section~\ref{sec-targets}.
    Lower panels: $\mathrm{position}-\Delta E$ spectra measured with the Split-Pole spectrometer placed at $85^\circ$. 
    Regarding $\alpha$-scattering, in both $T_3$ and Split-Pole plots, 
    only the elastic peak is clearly visible for $\noyA{Sm}{144}$, 
    while for $\noyA{Sm}{148}$, two inelastic peaks are also distinct.
    The first one is due to inelastic scattering to the first excited state of $\noyA{Sm}{148}$, the $2^+$ state at 550~keV.
    The second one, observed only for the run shown in this figure, can be attributed to inelastic scattering to the $4^+$ state at 1180~keV, with a possible contribution from the nearby $3^-$ state at 1162~keV.
   }
\end{figure}

To complete this setup, three silicon-detector telescopes (named here $T_1$, $T_2$, and $T_3$) were installed inside the scattering chamber to cover a broad range of laboratory angles, from $80^\circ$ to $163^\circ$, thus providing a  measurement of the elastic scattering cross section at backward angles. 
This angular region is crucial to constrain AOMP properties, since it is where the elastic scattering cross section is most sensitive to nuclear interactions.
To perform particle identification, each telescope consisted of a thin $\Delta E$ detector of $80~\mu m$ for energy-loss measurement and a thick $E$ detector of $500~\mu$m or $1000~\mu$m for total energy measurement. A 2-mm-diameter collimator was placed in front of each telescope, providing a compromise between angular resolution and counting statistics.
Examples of identification spectra obtained with one of the telescopes and the Split-Pole spectrometer are shown in Figure~\ref{fig:e-de_plot} for $\noyA{Sm}{144}$ and $\noyA{Sm}{148}$.

Using this hybrid setup, the total angular coverage ranged from $22^\circ$ to $163^\circ$.
To obtain a consistent data set between the Split-Pole spectrometer and the silicon array, consistency checks were performed with the two setups at 80° and 85°.  

The incident beam current was continuously monitored during data acquisition using a Faraday cup positioned at $0^\circ$, downstream of the target ladder. The cup was electrically isolated and connected to a charge integrator system. The integrated charge was recorded in parallel with event data and used to select the runs to be included in the analysis, according to beam-stability criteria assessed over time, both from run to run and within individual runs.

To minimize systematic uncertainties in the comparison between isotopes, each pair of measurements at a given angle was performed consecutively: the $^{144}$Sm and $^{148}$Sm targets were placed on the beam spot one after the other for each angular configuration. This approach ensured that both measurements experienced nearly identical beam and detector conditions, including beam energy, current, electronics response, and acquisition settings. 

\section{Samarium target characterization}
\label{sec-targets}
For this experiment, highly enriched samarium targets
were fabricated at the SIDONIE isotope separator facility of IJCLab ~\cite{CAMPLAN197037}. SIDONIE is a $40\ \text{kV}$ electromagnetic mass separator, 
based on a Nier–Bernas ion source~\cite{10.1063/1.1741192}, capable of delivering rare isotopes with high isotopic purity ($>99.9\%$) and high fluence (exceeding $10^{17}\ \text{atoms}/\text{cm}^2$ per day) via low-energy ($\sim 150\ \text{eV}$) ion beam deposition or implantation.
The deposition was controlled to achieve uniform layers by scanning the ion beam over the backing substrates. The enriched samarium isotopes were ionized in the Nier–Bernas source, separated by mass, then decelerated to approximately $150\ \text{eV}$ and deposited.

The two targets used in the experiment
were prepared by ion-beam deposition onto thin carbon backings of $50~\mu\text{g}/\text{cm}^2$, resulting in the following samarium surface densities:
\begin{itemize}
    \item $\noyA{Sm}{144}$ target: $44 \pm 4~\mu\text{g}/\text{cm}^2$;
    \item $\noyA{Sm}{148}$ target: $74 \pm 6~\mu\text{g}/\text{cm}^2$.
\end{itemize}

These targets have been characterized after the data acquisition by Rutherford Backscattering Spectrometry (RBS), to determine their effective thickness, spatial homogeneity and presence of contaminants.
The RBS measurements were performed with the ARAMIS accelerator of IJCLab~\cite{aramis}.
A $1.4$~MeV beam of $\alpha$ particles was directed at each target and the backscattered particles were detected 
at $165 ^\circ$ by a silicon detector.

\begin{figure}[htbp]
    \centering
    \includegraphics[width=1.0\textwidth]{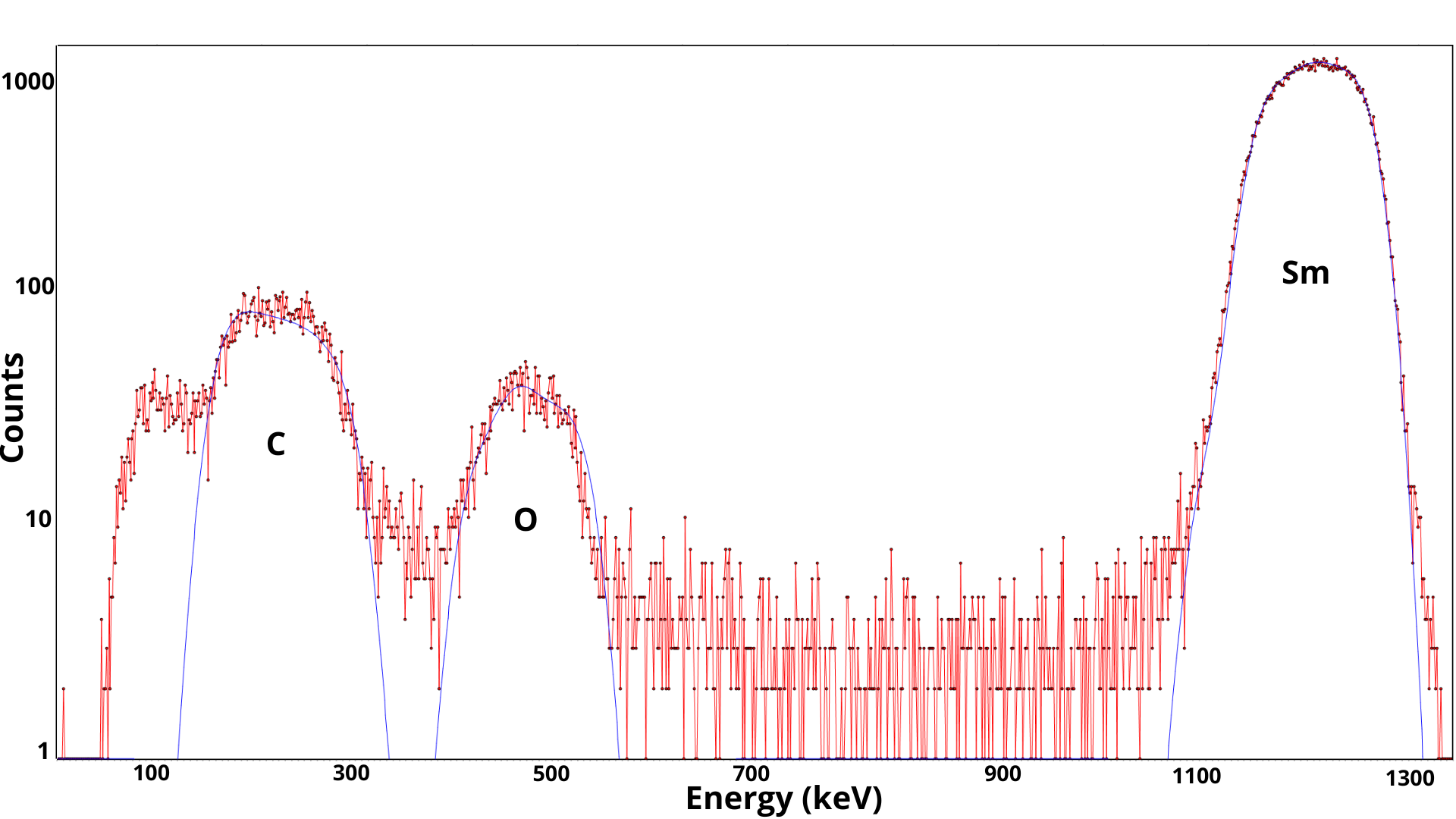}
    \caption{\label{fig:rbs_spectra}
    RBS spectrum for the central point of the $\noyA{Sm}{148}$ target. Carbon (C), oxygen (O) and samarium (Sm) peaks are indicated.
    The red line corresponds to experimental data, 
    and the blue line to the simulated spectrum fitted by SIMNRA.} 
\end{figure}

The backscattered spectra were analyzed using the SIMNRA software~\cite{Mayer-SIMNRA-1997} to extract surface densities. Figure~\ref{fig:rbs_spectra} displays a representative RBS spectrum fitted with SIMNRA. Three distinct peaks are clearly identified: a samarium peak, a carbon peak corresponding to the backing, and an oxygen peak arising from target oxidation. This oxygen contamination did not impact the analysis given the large difference in kinematics between elastic scattering on oxygen and samarium, suppressing any ambiguity in the peak identification.

The variations between the thickest and thinnest regions reached $20\%$. However, given the typical beam profile during the main experiment and the impact area verified by electronic microscopy, the uncertainty on the target thickness is limited to $7\%$.

\section{Data Analysis}
\label{sec-analysis}
 
The data collected during the experiment were processed using the ROOT data analysis framework \cite{root}. Energy calibration was performed using a combination of pulse generators and a standard triple $\alpha$ source ($^{241}$Am, $^{239}$Pu, and $^{244}$Cm). The pulse generators were used to check the linearity and electronic stability of each acquisition channel, while the $\alpha$ source provided absolute calibration points corresponding to known $\alpha$ energies. 
The energy resolution of the $\Delta E$ stage was measured to be about 80~keV (FWHM), while the $E$ detectors achieved a resolution of 20~keV (FWHM).
The triple $\alpha$ source was also used to determine the solid angle of each telescope independently.

For each angle, the elastic $\alpha$-scattering peak was clearly resolved from background and non-elastic events in the identification plots,
as shown in Figure~\ref{fig:e-de_plot}. 
In the case of $\noyA{Sm}{148}$, the inelastic peak on the first excited level at 550 keV could also be measured. 
For $\noyA{Sm}{144}$, the first excited state is at 1660 keV and does not appear 
in the spectra.

The absolute angular distributions obtained differed from 
literature values
by an important factor, 
indicating systematics that are thought to arise mainly from the determination of the beam intensity, biased by beam-tuning issues. 
For this reason, we focused on the extraction of relative observables.
{{We have checked that the correcting factor is constant along the experiment within the error bars. Furthermore, the study of the secondary reaction $\noyA{C}{12}(\alpha,p)$, occurring on the carbon backing, confirmed that this factor can be applied to both targets, then canceling for relative cross sections.}}

To compare the two samarium isotopes, 
the ratio between their elastic differential cross sections
is a particularly relevant observable, defined as:
\begin{equation}
\label{eq-elratio}
\elratio{Sm}{148}{144}
= \frac
{d\sigma/d\Omega (\noyA{Sm}{148}(\alpha,\alpha)\noyA{Sm}{148})}
{d\sigma/d\Omega (\noyA{Sm}{144}(\alpha,\alpha)\noyA{Sm}{144})}
\end{equation}
We could obtain this ratio in the angular range $[31^\circ-\,145^\circ]$.
The resulting values are given in Table~\ref{tab-data-el}. 
The total uncertainties were obtained as the quadratic sum of 
statistical uncertainties 
{{(limited to $3\%$ below $90^\circ$, and up to $27\%$ at backward angles)}}
and systematic uncertainties due to target thickness (up to $7\%$).
For two of the angles, values of {$\elratio{Sm}{148}{144}$} have been obtained both with Split-Pole and the $T_3$ telescope: 
the value given in the table is then the 
{{error-weighted average of the two values}}.
Taken separately for comparison,
at $\theta_{lab}=80^\circ$, we have {$\elratio{Sm}{148}{144}=0.995(60)$} for Split-Pole and $0.965(48)$ for $T_3$; 
at $\theta_{lab}=85^\circ$, we have {$\elratio{Sm}{148}{144}=0.990(24)$} for Split-Pole and $0.972(32)$ for $T_3$. 
In both cases, the values obtained with the two detection systems are compatible.

For $\noyA{Sm}{148}$, we have also studied the contribution of inelastic scattering to the first excited state. The results are given relatively to elastic scattering, thus canceling part of the experimental uncertainties. The corresponding observable is defined as:
\begin{equation}
\label{eq-inelratio}
\inelratio{Sm}{148}
= \frac
{d\sigma/d\Omega (\noyA{Sm}{148}(\alpha,\alpha_1)\noyA{Sm}{148})}
{d\sigma/d\Omega (\noyA{Sm}{148}(\alpha,\alpha)\noyA{Sm}{148})}    
\end{equation} 
where $\alpha_1$ is the ejectile after inelastic scattering producing the first exited state of $\noyA{Sm}{148}$.
The value of $\inelratio{Sm}{148}$ is not impacted by fluctuations in beam intensity and target thickness: as a consequence, 
experimental uncertainty is much lower than for $\elratio{Sm}{148}{144}$, 
and data could be obtained in the angular range of $[80^\circ-\,163^\circ]$. 
Below $80^\circ$, the inelastic peak cannot be distinguished from the tail of the elastic peak, due to the much higher elastic cross section at forward angles.
The corresponding results are given in Table~\ref{tab-data-inel}. 
The value at $80^\circ$ is only from Split-Pole, since for the silicon detector telescope $T_3$ the inelastic peak could not be clearly defined at this angle. At $85^\circ$, the statistics from Split-Pole and from $T_3$ were summed. 
{{Taken separately, the values obtained with both detectors at $85^\circ$ present a slight discrepancy: 
$\inelratio{Sm}{148}=0.0248(12)$ for Split-Pole and $0.0300(19)$ for $T_3$.}}

The data tables give the angle $\theta_{lab}$ in the laboratory frame. In the next section, the figures will display curves as a function of the angle 
$\theta_{cm}$ in the center-of-mass frame, except when representing isotopic ratios which are drawn as a function of $\theta_{lab}$ 
(since for a given $\theta_{lab}$ fixed by the experimental setup, $\theta_{cm}$ is slightly different for the two isotopes).
 
\begin{table}
\caption{Experimental data obtained in the present work for 
the elastic differential cross section ratio between $\noyA{Sm}{148}$ and $\noyA{Sm}{144}$,
with associated uncertainties. 
All laboratory frame scattering angles include an uncertainty of $\pm~1 ^\circ$, originating from the angular reading accuracy and from the angular acceptance. 
\label{tab-data-el}}
\begin{center}
\begin{tabular}{cc}
$\theta_{lab}$ (deg) & \multicolumn{1}{c}{\quad $\elratio{Sm}{148}{144}$} \\ 
\hline
31	&	1.10(12)	\\
41	&	1.06(8)	\\
50	&	1.08(10)	\\
60	&	1.14(11)	\\
70	&	1.03(11)	\\
80	&	{{0.977(37)}}	\\
85	&	{{0.984(19)}}	\\
97	&	0.904(101)	\\
103	&	0.887(100)	\\
108	&	1.15(22)	\\
109	&	0.988(119)	\\
111	&	0.885(110)	\\
112	&	0.943(189)	\\
116	&	1.06(16)	\\
121	&	0.955(173)	\\
133	&	0.904(110)	\\
135	&	1.02(20)	\\
139	&	0.836(105)	\\
144	&	0.982(292)	\\
145	&	0.819(126)	\\
\end{tabular}
\end{center}
\end{table}

\begin{table}
\caption{Experimental data obtained in the present work for 
differential cross section ratio between inelastic scattering 
to the first excited state of $\noyA{Sm}{148}$ and 
elastic scattering on the same nucleus,
with associated uncertainties. 
Uncertainties for $\theta_{lab}$ are the same as in table~\ref{tab-data-el}.
\label{tab-data-inel}}
\begin{center}
\begin{tabular}{cc}
$\theta_{lab}$ (deg) & \multicolumn{1}{c}{\quad $\inelratio{Sm}{148}$} \\ 
\hline
80	&	0.0180(23)	\\
85	&	{{0.0263(10)}}	\\
93	&	0.0477(22)	\\
97	&	0.0458(24)	\\
99	&	0.0551(22)	\\
103	&	0.0547(31)	\\
105	&	0.0625(30)	\\
108	&	0.0530(61)	\\
109	&	0.0684(78)	\\
112	&	0.0794(74)	\\
115	&	0.0864(41)	\\
121	&	0.0907(44)	\\
127	&	0.103(9)	\\
133	&	0.118(10)	\\
139	&	0.138(12)	\\
144	&	0.150(20)	\\
145	&	0.135(24)	\\
151	&	0.152(11)	\\
157	&	0.145(12)	\\
163 &	0.156(17)	\\
\end{tabular}
\end{center}
\end{table}


\section{Discussion}
\label{sec-discussion}

Let us now compare experimental results with model predictions in order to determine to what extent the {\AOMP} can be characterized by these observables. A typical study would be to fit the elastic angular distributions, in order to define for the first time a local $\alpha$ optical potential for $\noyA{Sm}{148}$ 
and compare it to the well-studied case of $\noyA{Sm}{144}$, 
but a fitting procedure is not significative here due to the large uncertainties on our data. 
However, interesting features can be discussed by studying the isotopic ratio of elastic scattering differential cross sections. 
We also investigate the model dependence of the inelastic scattering to the first excited level of $\noyA{Sm}{148}$. 
Finally, we will discuss the role of AOMP in low-energy
$\alpha$-induced reactions that impact astrophysical rates involved in
the $p$-process.

\subsection{Alpha optical model potentials}

We first present the {\AOMP} models that have been used in this study. Our model calculations have been performed using the TALYS code, version 2.0~\cite{talys}, which includes eight options for the choice of the {\AOMP} model. 
Further, we have integrated in TALYS-2.0 the Atomki-V2 {\AOMP} defined by Mohr \textit{et al.} in the supplement of~\cite{mohr_2020}, 
adapting the guidance offered in~\cite{aomp9}. 
In addition, we have defined  {\AOMP} models tuned for the study of our two samarium isotopes, 
by fitting $\noyA{Sm}{144}(\alpha,\alpha)\noyA{Sm}{144}$ data by Mohr-1997~\cite{Mohr-97} 
and extending the results to $\noyA{Sm}{148}$ \textit{via} constraints that will be specified.

The optical model describes the scattering and absorption of a projectile by a nucleus with the help of a complex potential. When the projectile is an $\alpha$ particle (spinless), the nuclear part of this potential can be written, as a function of the distance $r$: 
\begin{equation}
\mathcal{U}(r)=
-\mathcal{V}(r)-i\mathcal{W}(r)
\end{equation}
The terms are often described using Woods-Saxon form factors involving two geometrical parameters, radius $R$ and diffuseness $a$:
\begin{equation}
f(r,R,a) = \left[1+\exp\frac{r-R}{a}\right]^{-1}    
\end{equation} 
For most existing models, the imaginary part is described by a volume term (Woods-Saxon) sometimes supplemented by a surface term (Woods-Saxon derivative), with each term involving depth, radius and diffuseness parameters. 
The geometrical parameters are then distinguished by an index {{$i=V,S$}} to identify the corresponding terms.
For a nucleus of mass number $A$, the reduced radius is $r_i=R_i/A^{1/3}$ and the imaginary part of the potential reads:
\begin{eqnarray}
\mathcal{W}(r)&=&
W_V \times f(r,r_VA^{1/3},a_V)\nonumber\\
&+& {W_S \times (-4a_S) \times \frac{d}{dr}f(r,r_S A^{1/3},a_S)}
\end{eqnarray}
Concerning the real part, models split in two families: some of them use the phenomenological Woods-Saxon description ({\WS}), others use the microscopic double-folding approach ({\DF}). 
For {\WS} models, the real potential reads: 
\begin{equation}
\mathcal{V}(r)=V\times f(r,r_vA^{1/3},a_v)
\end{equation}
For {\DF} models, the real potential is described by a double-folding procedure, using density profiles and a density-dependent nuclear interaction (see e.g.~\cite{aomp345}).

\begin{table}
\caption{Eight {\AOMP} models available in TALYS-2.0 code, plus Atomki-V2 potential integrated as a ninth option (\aomp{9}). Column 3 characterises the real part as Woods-Saxon ({\WS}) or double-folding ({\DF}). Column 4 specifies if the imaginary part is described by a volume term (V) or volume+surface terms (V+S).
\label{tab-talys-aomp}}
\begin{tabular}{ccccc}
Model & Reference & Re. part & Im. part & Comment \\ 
\hline
\aomp{1} & Koning-Delaroche-2003~\cite{KONING2003231} & \WS & V+S & From nucleon-nucleus OMP \\ 
\aomp{2} & McFadden-Satchler-1966~\cite{aomp2} & \WS & V & Historical {\AOMP} \\ 
\aomp{3} & Demetriou-2002~\cite{aomp345} & {\DF} & V &  \\ 
\aomp{4} & Demetriou-2002~\cite{aomp345} & {\DF} & V+S &  \\ 
\aomp{5} & Demetriou-2002~\cite{aomp345} & {\DF} & V+S & Dispersive \\ 
\aomp{6} & Avrigeanu-2014~\cite{aomp6} & \WS & V+S & Default option in TALYS \\ 
\aomp{7} & Nolte-1987~\cite{aomp7} & \WS & V & Fitted to 30 MeV data \\ 
\aomp{8} & Avrigeanu-1994~\cite{aomp8} & \WS & V & \aomp{7} with energy dependence \\ 
\aomp{9} & Mohr-2020~\cite{mohr_2020} & {\DF} & V & Added to TALYS \\ 
\end{tabular}
\end{table}

\begin{figure}
\begin{center}
\includegraphics[width=1.0\textwidth]{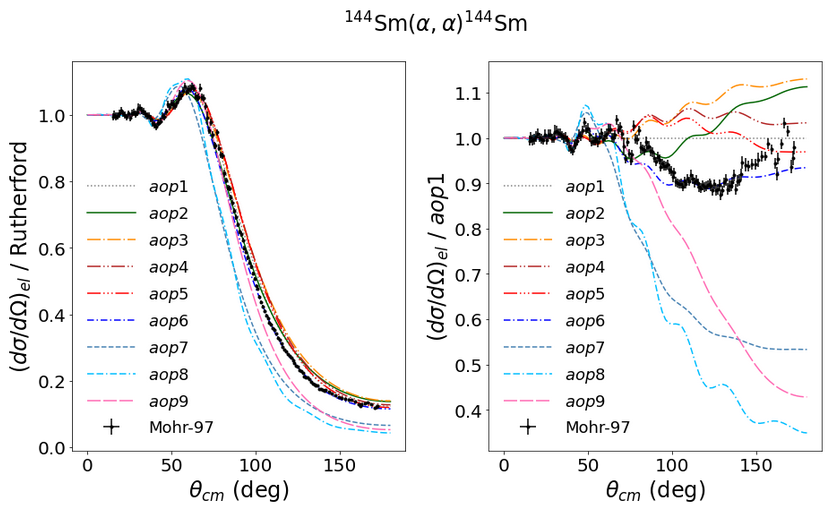}
\caption{Elastic angular distribution: experimental data by Mohr \textit{et al.}~\cite{Mohr-97} compared to TALYS results with different standard {\AOMP} models.
Left: normalization to Rutherford cross section.
Right: normalization to \aomp{1} to emphasize model dependence.}
\label{fig-CSR}
\end{center}
\end{figure}

The {\AOMP} used in this work are listed in Table~\ref{tab-talys-aomp}.
Several of these models have been designed 
as an input in nucleosynthesis reaction networks, 
and include constraints from experimental data on $\noyA{Sm}{144}+\alpha$ elastic scattering by Mohr \textit{et al.}~\cite{Mohr-97} and radiative capture by Somorjai \textit{et al.}~\cite{Somorjai-97}, thus we can expect them to describe correctly the elastic angular distribution of $\noyA{Sm}{144}(\alpha,\alpha)\noyA{Sm}{144}$. 
We can see in Figure~\ref{fig-CSR} that indeed, 
although none of these global models reproduces accurately the experimental data, 
the Avrigeanu-2014 potential (\aomp{6}) gives very close results, 
and the curves from the Demetriou-2002 series of potentials (\aomp{3}, \aomp{4}, \aomp{5}) are only slightly above at backward angles.
We can note however that the historical McFadden-Satchler potential (\aomp{2}) gives results similar to the modern potentials by Demetriou-2002, while the global Atomki-V2 from Mohr-2020 (\aomp{9}) strongly underestimates the data at backward angles.

Let us now introduce our adapted potentials. Local potentials for $\noyA{Sm}{144}(\alpha,\alpha)$ have been fitted in previous work by Mohr-1997~\cite{Mohr-97} and Kiss-2022~\cite{Kiss-22}. In the present work, however, we want to investigate the role of potential shape in the comparison between $\alpha$ elastic scattering on $\noyA{Sm}{144}$ and $\noyA{Sm}{148}$. For this, we work on potentials involving a minimal number of parameters that can be reasonably adapted from $\noyA{Sm}{144}$ to $\noyA{Sm}{148}$:
{{{\aomp{2} (WS) and \aomp{9} (DF). In each case, two parameters are fitted to the Mohr-1997 data, and the corresponding potential for $\noyA{Sm}{148}$ is deduced from specific criteria that can be employed for these models, as will be detailed below.}}}
The parameters are fitted using a $\chi^2$ minimization procedure. 
{{The uncertainty given on the corresponding parameters is determined through the inverse of the $\chi^2$ curvature matrix in the space of the varied parameters, following the statistical analysis presented in~\cite{Fattoyev-11}. Here, this uncertainty is not propagated further, since this would not lead to significant variations of the curves and a full analysis of the theoretical uncertainties is beyond the scope of this work.}}

Our first adapted potential is based on the {\AOMP} by McFadden and Satchler, \aomp{2}~\cite{aomp2}.
This well-known model was a pioneering attempt to obtain an {\AOMP} that can be used for a wide range of nuclei, where both real and imaginary parts are described by a Woods-Saxon volume term with identical reduced radius $\rMS$ and diffuseness $\aMS$.
The corresponding parameters for $\noyA{Sm}{148}$ have been studied in the work by Badawy-1978~\cite{Badawy-78}, where excitation functions of $\noyA{Sm}{A}(\alpha,\alpha)\noyA{Sm}{A}$ (among others) are studied for different samarium isotopes at collision energies around the Coulomb barrier. With fixed values of the depths, $V=200$ MeV and $W=20$ MeV, it is shown in~\cite{Badawy-78} that the radius $R$ and diffuseness $a$ for $\noyA{Sm}{148}$ obey the following relationship:
\begin{equation}
R + \alpha \times a = 11.19\pm 0.03 \;\text{fm}
\label{eq-Badawy}    
\end{equation}
with $\alpha \simeq 7$ (common to the various nuclei under study).

We have first used the data from Mohr-1997~\cite{Mohr-97}
to fit \aomp{2} radius and diffuseness parameters for $\noyA{Sm}{144}$, adopting the depth values used in~\cite{Badawy-78}.
We call the resulting potential \MSM. Then, we have used Eq.~(\ref{eq-Badawy}) to define three {\AOMP} versions for $\noyA{Sm}{148}$, called {\MSBa}, {\MSBb} and {\MSBc}. The corresponding parameters are summerized in Table~\ref{tab-aomp2-fit}.
Note that the default \aomp{2} parameters are identical for $\noyA{Sm}{144}$ and $\noyA{Sm}{148}$, while the inclusion of the Badawy constraint~(\ref{eq-Badawy}) leads to a larger spatial extension of the potential for $\noyA{Sm}{148}$:
\begin{itemize}
\item For \MSBa, this corresponds to a larger diffuseness (with a reduced radius taken from \MSM).
\item For \MSBc, this corresponds to a larger radius (with a diffuseness taken from \MSM).
\item {\MSBb} corresponds to an average of the two above choices.
\end{itemize}

\begin{table}
\caption{
Parameters of different versions of McFadden and Satchler potentials used in this work (see text for details). For samarium isotopes, \aomp{2} is the default version while \MSM, -Ba, -Bb, -Bc are our adapted versions with depths $V$ and $W$ fixed to the values adopted in~\cite{Badawy-78}. Best fit parameters from~\cite{Tabor-76} for neodymium isotopes are also given (smaller depth in this case is associated with larger radius). 
\label{tab-aomp2-fit}}
\begin{tabular}{cccccc} 
Potential name & Nucleus & $V$ (MeV) & $W$ (MeV) & $\rMS$ (fm)  & $\aMS$ (fm) \\ 
\hline
\aomp{2} & $\noyA{Sm}{144}$ & 185 & 25 & 1.4 & 0.52 \\ 
\aomp{2} & $\noyA{Sm}{148}$ & 185 & 25 & 1.4 & 0.52 \\ 
\MSM & $\noyA{Sm}{144}$ & 200 & 20 & {{1.358(4)}} & {{0.558(1)}} \\ 
\MSBa & $\noyA{Sm}{148}$ & 200 & 20 & {{1.358}} & {{0.572}} \\
\MSBb & $\noyA{Sm}{148}$ & 200 & 20 & {{1.368}} & {{0.565}} \\
\MSBc & $\noyA{Sm}{148}$ & 200 & 20 & {{1.377}} & {{0.558}} \\
\MSTNda & $\noyA{Nd}{142}$ & 100 & 10 & 1.462 & 0.543 \\
\MSTNdb & $\noyA{Nd}{146}$ & 100 & 10 & 1.457 & 0.559 \\
\end{tabular} 
\end{table}

We have also used the Mohr-1997 data to fit the Atomki-V2 potential \aomp{9},
by adjusting the depth and width of the real potential given in~\cite{aomp9}, which is calculated with a double-folding procedure. Note that for this {\AOMP}, the imaginary part is fixed to a deep and sharp volume term that is identical for all nuclei, and we did not modify it. 
The real part for the default version is fixed by the double-folding procedure and a prescription for the real-potential volume integral $J_R$. 
This integral, defined as 
$J_R=\int{\mathcal{V}(r)d^3r}/(A_P A)$ (where $A_P$ and $A$ are the projectile and target mass numbers), 
is known as an important quantity to characterize the optical potential, with weak energy dependence.
For Atomki-V2, the real potential depth is scaled in order to have 
$J_R=342.4$ MeV.fm$^3$ for semi-magic nuclei 
and $J_R=371.0$ MeV.fm$^3$ for non-magic nuclei.
In our adapted version, named \AtoM, the fit parameters are the corrective factors to the real potential: $\lambda$ (depth parameter) and $w$ (width parameter), such that for 
potential $V_f(r)$ obtained by double-folding procedure, 
the final real potential is (see e.g.~\cite{Mohr-97}):
\begin{equation}
\mathcal{V}(r)=\lambda \times V_f(r/w)
\end{equation}
The $\chi^2$ minimization leads to 
{{$\lambda=0.928(2)$ and $w=0.9802(2)$}}. The same parameters are adopted for $\noyA{Sm}{148}$, since the expected difference between $\noyA{Sm}{148}$ (non-magic) and $\noyA{Sm}{144}$ (magic) is included in the original potential files given in~\cite{aomp9} through the scaling of the volume integral $J_R$.

\begin{figure}
\begin{center}
\includegraphics[width=1.0\textwidth]{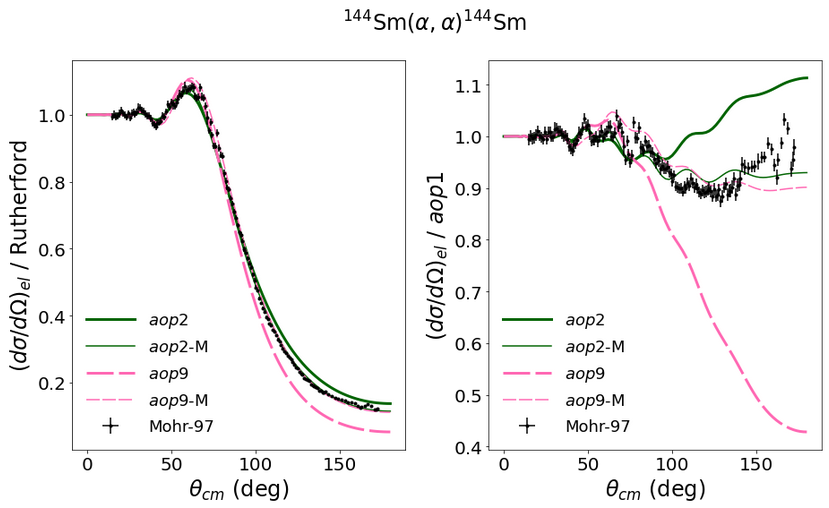}
\caption{
Similar to Figure~\ref{fig-CSR}, with standard {\AOMP} models \aomp{2}, \aomp{9} compared to potentials fitted to the Mohr-1997 experimental data~\cite{Mohr-97} (see text).}
\label{fig-CSR-fit}
\end{center}
\end{figure}

The elastic angular distributions obtained with our adapted potentials, compared to their original versions, are shown in Figure~\ref{fig-CSR-fit}.

\subsection{Elastic isotopic ratio}

\begin{figure}
\begin{center}
\includegraphics[width=1.0\textwidth]{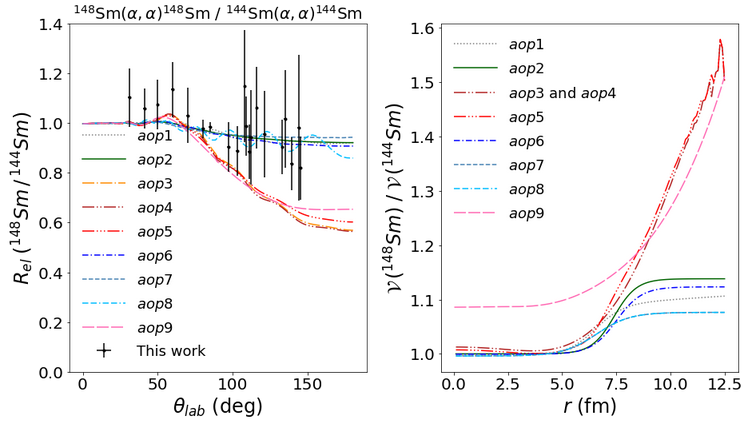}
\caption{Left: TALYS prediction for the elastic cross-section ratio $\elratio{Sm}{148}{144}$ for the nine standard {\AOMP} models, compared with experimental data from this work. 
{\WS} models: \aomp{1}, 2, 6, 7, 8.
{\DF} models: \aomp{3}, 4, 5, 9.
Right: real potential ratio $\mathcal{V}(\noyA{Sm}{148})/\mathcal{V}(\noyA{Sm}{144})$ as a function of the distance $r$ for the nine standard {\AOMP} models
(the two Demetriou potentials \aomp{3} and \aomp{4} differ only by the imaginary part, hence they correspond here to the same curve).
}
\label{fig-elratio-all}
\end{center}
\end{figure}

The elastic isotopic ratio of $\noyA{Sm}{148}$ and $\noyA{Sm}{144}$, 
denoted $\elratio{Sm}{148}{144}$, is defined by Eq.~(\ref{eq-elratio}).
This observable is shown in the left panel of Figure~\ref{fig-elratio-all}, with theoretical curves corresponding to the nine TALYS options in their original version. A striking difference appears in the trends of the two {\AOMP} families, {\WS} and {\DF}. The different double-folding potentials have quite different shapes of the imaginary part, thus we can deduce that the observed common behavior reflects a property of the real part. A similar separation between {\WS} and {\DF} models appears in the ratio $\mathcal{V}(\noyA{Sm}{148})/\mathcal{V}(\noyA{Sm}{144})$ as a function of the distance $r$,
where the {\WS} models converge to a constant value beyond $r\sim 8$ fm, while the {\DF} models increase steeply:
see the right panel of Figure~\ref{fig-elratio-all}. We can notice the specific behavior of \aomp{9}, for which the potential ratio in the central region is sensitively larger than one; this reflects the Atomki-V2 construction, with the volume integral $J_R$ normalized to a different value for semi-magic and non-magic nuclei~\cite{aomp9}. However, this specificity has no impact on the elastic ratio, whose behavior is similar to other {\DF} potentials. Thus we can conclude that the real potential ratio in the  peripheric region is a determining feature for the ratio 
$\elratio{Sm}{148}{144}$.

Our experimental data for this observable are affected by large uncertainties. 
However, beyond 120$^\circ$, it presents a discrepancy at the 1 $\sigma$ level with the {\DF} models.
Although the data seem globally closer to the {\WS} family, the points at forward angle are sensitively above the expected ratio, which should be close to one, then revealing a systematic error toward upper values.
Another indication of such upward shift comes from the comparison with earlier experimental data on neodymium, as will be discussed afterward.

\begin{figure}
\begin{center}
\includegraphics[width=1.0\textwidth]{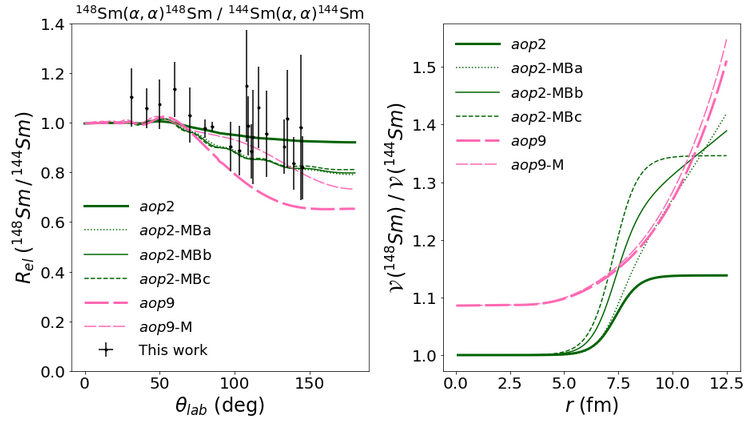}\caption{Similar to Figure~\ref{fig-elratio-all}, with {\AOMP} models fitted to $\noyA{Sm}{144}$ and $\noyA{Sm}{148}$ compared to their standard versions \aomp{2} and \aomp{9} (see text).}
\label{fig-elratio-fitted}
\end{center}
\end{figure}

We go a step further and perform a similar analysis with the potentials we built according to known constraints on $\noyA{Sm}{144}$ and $\noyA{Sm}{148}$.
We show in Figure~\ref{fig-elratio-fitted} the elastic and real potential ratios obtained with the fitted potentials 
\MSM, \MSBa, \MSBb, {\MSBc} and \AtoM, compared to their original versions \aomp{2} ({\WS} model) and \aomp{9} ({\DF} model). 
For the fitted versions of \aomp{2}, the potentials used for these ratios are labeled {\MSMBa}, {\MSMBb} and {\MSMBc}, indicating that {\MSM} is used for $\noyA{Sm}{144}$, and respectively {\MSBa}, {\MSBb} and {\MSBc} are used for $\noyA{Sm}{148}$. 
A striking feature emerges: the adapted potentials show a behavior that lay in between the two families, {\WS} and {\DF}.
The three modified versions of \aomp{2} show very similar elastic isotopic ratios. Their real potential isotopic ratios present a crossing around $r\simeq$ 11~fm, which is a direct consequence of the condition~(\ref{eq-Badawy}); see~\cite{Badawy-78} for details. On the other hand, the behavior observed for \aomp{9} and {\AtoM} cannot be explained so easily by the real potential ratio, for which the very slight rescaling of the radius variable applied for {\AtoM} (corrective factor $w=0.980$) causes very little shift except in the very peripheral region. As for the corrective factor $\lambda=0.93$ applied to the potential depth, it is canceled in this ratio.

\begin{figure}
\begin{center}
\includegraphics[scale=0.55]{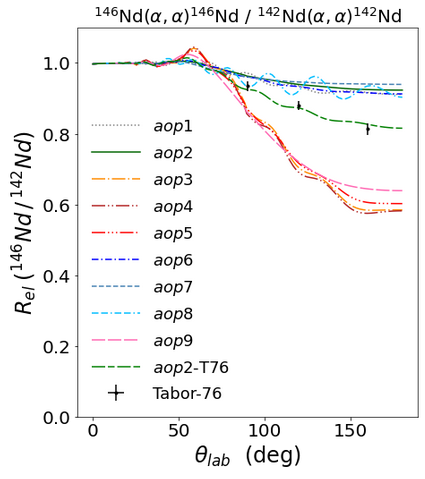}
\caption{
TALYS prediction for the elastic cross-section ratio $\elratio{Nd}{146}{142}$ for the nine standard {\AOMP} models, compared with experimental data (Tabor-76) and fitted {\AOMP} from~\cite{Tabor-76} for neodymium isotopes $\noyA{Nd}{142}$ and $\noyA{Nd}{146}$ ($aop2$-T76).
}
\label{fig-elratio-Nd}
\end{center}
\end{figure}

Although our experimental uncertainties do not allow to constrain the {\AOMP} models, we can reasonably expect that similar data obtained in better experimental conditions would be valuable to progress in the shaping of a global $\alpha$ potential. Such global potential has to take into account the isotopic dependence and especially the effect of shell closures. 
To illustrate this point, we show in Figure~\ref{fig-elratio-Nd} similar curves obtained for $\noyA{Nd}{146}$ and $\noyA{Nd}{142}$, a system analogous to the $\noyA{Sm}{144}-\noyA{Sm}{148}$ pair, with two protons less. The data from Tabor \textit{et al.}~\cite{Tabor-76} (where error bars correspond to statistical uncertainties) exclude the two extreme behaviors shown by both standard {\WS} and {\DF} models, while the fitted {\AOMP} determined in~\cite{Tabor-76} leads to an intermediate curve, similar to the results we have obtained with samarium isotopes. The fitted potential is the one giving the best fit in~\cite{Tabor-76}, whose parameters are listed in Table~\ref{tab-aomp2-fit}.

\subsection{Inelastic scattering on \texorpdfstring{$\noyA{Sm}{148}$}{148Sm}}

The first excited level of $\noyA{Sm}{148}$ is a $2^+$ state at energy $E_1=550$ keV, much below the equivalent one in the spherical nucleus $\noyA{Sm}{144}$, at 1660 keV. As a consequence, the $\alpha$ inelastic scattering to the first level of $\noyA{Sm}{148}$ is observed in our experimental data. 
We have measured its ratio to elastic scattering: 
this observable, denoted $\inelratio{Sm}{148}$, is defined by Eq.~(\ref{eq-inelratio}).

Calculations of inelastic differential cross sections can be performed with TALYS, where the ECIS code~\cite{ECIS} is used as a subroutine. Since $\noyA{Sm}{148}$ has a weak deformation (with experimental value $\beta_2=0.142~\pm~0.003$~\cite{NNDC}), it is treated by default as a spherical nucleus in TALYS, with cross sections calculated in the Distorted Wave Born Approximation (DWBA) regime. However, the coupling scheme can be modified in the TALYS structure files in order to perform coupled-channel (CC) calculations with different settings. 
{{For the inelastic scattering observables, the role of the AOMP will be discussed more briefly, focusing on the following Woods-Saxon potentials: 
(i) \aomp{2}, the historic McFadden and Satchler potential, and its modified versions adapted to $\noyA{Sm}{148}$;
(ii) \aomp{6}, the modern potential of Avrigeanu \textit{et al.}, which is the global potential that gives the best description of elastic data $\noyA{Sm}{144}(\alpha,\alpha)$.}}

\begin{figure}
\begin{center}
\includegraphics[width=1.0\textwidth]{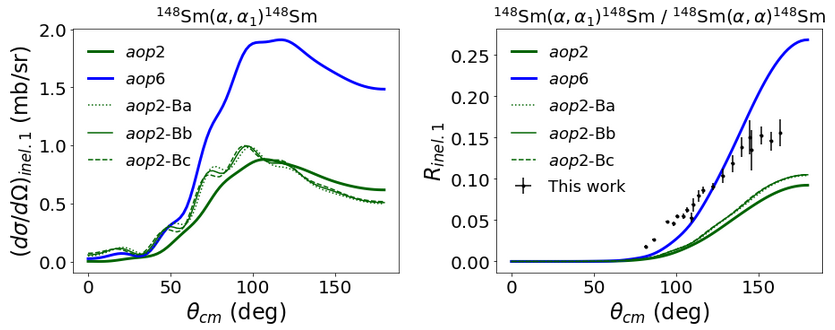}
\caption{Left: differential cross section of $\alpha$ inelastic scattering to the first level of $\noyA{Sm}{148}$ for different {\AOMP}, in the DWBA approach. Right: ratio to elastic differential cross section, with experimental data from this work.
}
\label{fig-inel1-pot}
\end{center}
\end{figure}

We show in Figure~\ref{fig-inel1-pot} the $\alpha$ inelastic scattering differential cross section to the first excited state of $\noyA{Sm}{148}$ obtained with different {\AOMP} models of {\WS} type, in the DWBA approach. The predicted ratio to elastic scattering is also plotted with our experimental data. It is seen that the result is very sensitive to the {\AOMP}. Further, the inelastic ratio is affected both by inelastic and elastic cross sections, with effects that can reinforce or compensate each-other; hence the variations in curve ordering when changing from inelastic cross section to inelastic over elastic ratio. 
One can also notice that the three versions of adapted potentials {\MSBa}, {\MSBb} and {\MSBc} lead to very similar results. In all cases, the theoretical curves are not able to reproduce the data (except for the Avrigeanu-2014 potential \aomp{6} around $\theta=125^o$).

\begin{figure}
\begin{center}
\includegraphics[width=1.0\textwidth]{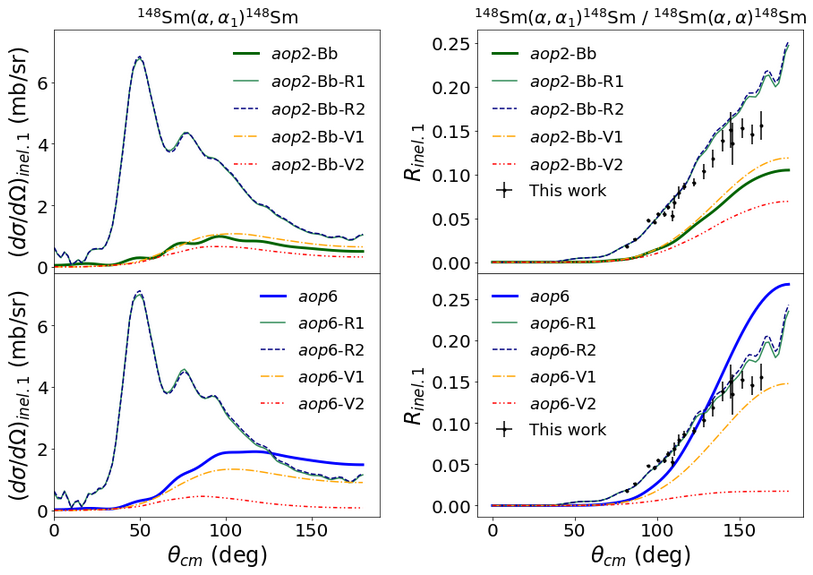}
    \caption{Impact of coupling scheme ($R1$, $R2$, $V1$, $V2$, see text) for two {\AOMP} (\aomp{6} and {\MSBb}). The DWBA curves are also shown (bold line). Left: differential cross section of $\alpha$ inelastic scattering on the first level of $\noyA{Sm}{148}$. Right: ratio to elastic differential cross section, with experimental data from this work.
}
\label{fig-inel1-CC}
\end{center}
\end{figure}

In Figure~\ref{fig-inel1-CC}, we explore the sensitivity of the inelastic prediction to the coupling scheme, using two {\AOMP} models: {\MSBb} and \aomp{6}. 
We compare DWBA results to CC calculations with four different coupling schemes, called $R1$, $R2$, $V1$ and $V2$. The letter indicates the type of excitation ($V$ = vibrational, $R$ = rotational) and the number indicates the number of coupled levels (1: $2^+$ state, 2: $2^+$ and $4^+$ states). Level energies are $E(2^+)=550$ keV and $E(4^+)=1180$ keV. 
Again, it is seen that the results are very sensitive to these different options. A rotational coupling strongly increases the inelastic cross section at medium angles with respect to DWBA and vibrational coupling, although this feature is hidden in the ratio to elastic scattering, which becomes significant only at backward angles. Due to its small deformation, $\noyA{Sm}{148}$ was previously treated in the literature as vibrational~\cite{OBIAJLINWA1989341}. However, its level scheme~\cite{NNDC} rather shows the $2^+$ and $4^+$ states as members of a rotational band, although it clearly differs from a pure rotor. Further, extending the coupling to higher levels leads to a good convergence in the rotational scheme, not in the vibrational one. The rotational coupling gives the best reproduction of the experimental data, and hinders the sensitivity to the {\AOMP} model.

\begin{figure}
\begin{center}
\includegraphics[width=1.0\textwidth]{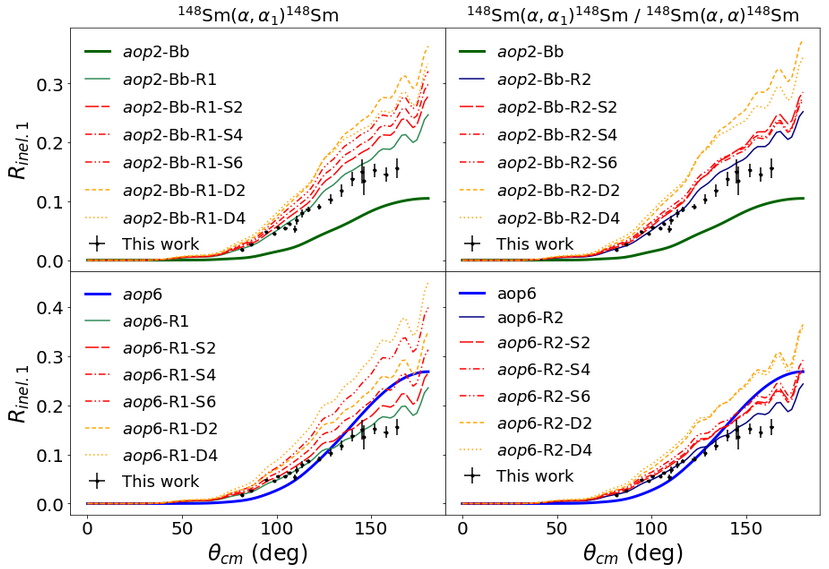}
\caption{
Inelastic ratio: impact of the deformation parameter used in rotational coupling schemes, for two {\AOMP} (\aomp{6} and {\MSBb}). 
Left: one coupled level ($R1$). Right: two coupled levels ($R2$).
The DWBA curves are also shown (bold line).
The thin full lines correspond to rotational coupling with experimental $\beta_2$ value.
For the other curves, the extensions
S$n$ (resp. D$n$) indicate deformations calculated with SHFB (resp. FRDM) up to $\beta_n$.
}
\label{fig-inel1-beta}
\end{center}
\end{figure}

We further consider the impact of the deformation parameter on the CC results with rotational coupling, see Figure~\ref{fig-inel1-beta}.
The $\beta_2$ value is known experimentally~\cite{NNDC}, but it can also be given by theoretical models: we use here deformation values from Skyrme-Hartree-Fock-Bogoliubov (SHFB)~\cite{Goriely-09} and Finite Range Droplet Model (FRDM)~\cite{Moller-16} approaches. These calculations give $\beta_\lambda$ parameters of several multipolarities: $\beta_2$, $\beta_4$ and $\beta_6$. 
When the deformation parameters are given in the coupling scheme up to $\beta_n$, 
the corresponding versions are called SHFB-$\beta_n$, FRMD-$\beta_n$.
In the figure legend, these notations are abbreviated as S$n$ (Skyrme) and D$n$ (Droplet).
The different deformation values are listed in Table~\ref{tab-def}.
As noticed earlier, as soon as a rotational coupling is switched on, the dependence on the {\AOMP} is washed out. However, the curves have a sensitive dependence on the deformation parameters. For $R1$ models, all versions of deformation give distinct curves. For $R2$ models however, we can observe a convergence for different multipole orders within a given model. In any case, our data is best described by the rotational coupling scheme with the experimental $\beta_2$ value.

\begin{table}
\caption{
Deformation parameters $\beta_\lambda$ from experimental measurement~\cite{NNDC} and theoretical calculations (SHFB~\cite{Goriely-09}, GHFB~\cite{Goriely-16}, FRDM~\cite{Moller-16}). Gogny-HFB (GHFB) is given for completeness, but its $\beta_2$ value is very close to SHFB and the corresponding curves for GHFB-$\beta_2$ and SHFB-$\beta_2$ are almost similar.
\label{tab-def}}
\begin{center}
\begin{tabular}{cccc} 
Source & $\beta_2$ & $\beta_4$ & $\beta_6$ \\ 
\hline
Exp. & 0.142(3) &   &  \\ 
SHFB & 0.151 & 0.04 & 0.056 \\ 
GHFB & 0.15 & 0 & 0 \\ 
FRDM & 0.172 & 0.06 & 0 \\ 
\end{tabular} 
\end{center}
\end{table}

\subsection{Impact of optical potential on \texorpdfstring{$\alpha$}{alpha}-induced reactions}

\subsubsection{Impact on \texorpdfstring{$(\alpha,\gamma)$}{(a,g)} reaction at astrophysical energy}

The $\alpha$-nucleus potential is a key ingredient for the numerous $(\gamma,\alpha)$ photodisintegrations that occur in the $\gamma$-process. The {\AOMP} for $\noyA{Sm}{144}$ and $\noyA{Sm}{148}$ are thus critical, for the $\noyA{Gd}{148}(\gamma,\alpha)\noyA{Sm}{144}$ and $\noyA{Gd}{152}(\gamma,\alpha)\noyA{Sm}{148}$ rates, respectively. According to~\cite{Rauscher-13} (see Figure 12 of this reference), $\noyA{Gd}{148}$ and  $\noyA{Gd}{152}$ correspond to competition points between different reactions, so the $(\gamma,\alpha)$ rate at these points is crucial for the direction of the flux in the reaction network. Besides, it is known that $(\alpha,\gamma)$ radiative capture cross-sections are the best-suited laboratory observables to determine the $(\gamma,\alpha)$ astrophysical reaction rates (see e.g.~\cite{Mohr-07}). 

We investigate in this section the impact of {\AOMP} determination through $\alpha$ scattering close to Coulomb barrier on the $(\alpha,\gamma)$ cross section at astrophysical energy. We consider the reactions 
$\noyA{Sm}{144}(\alpha,\gamma)\noyA{Gd}{148}$ and 
$\noyA{Sm}{148}(\alpha,\gamma)\noyA{Gd}{152}$, using TALYS to calculate their cross-section in the relevant energy range (the Gamow window is typically between 5 and 12 MeV) with different versions of {\AOMP}. 

\begin{figure}
\begin{center}
\includegraphics[width=1.0\textwidth]{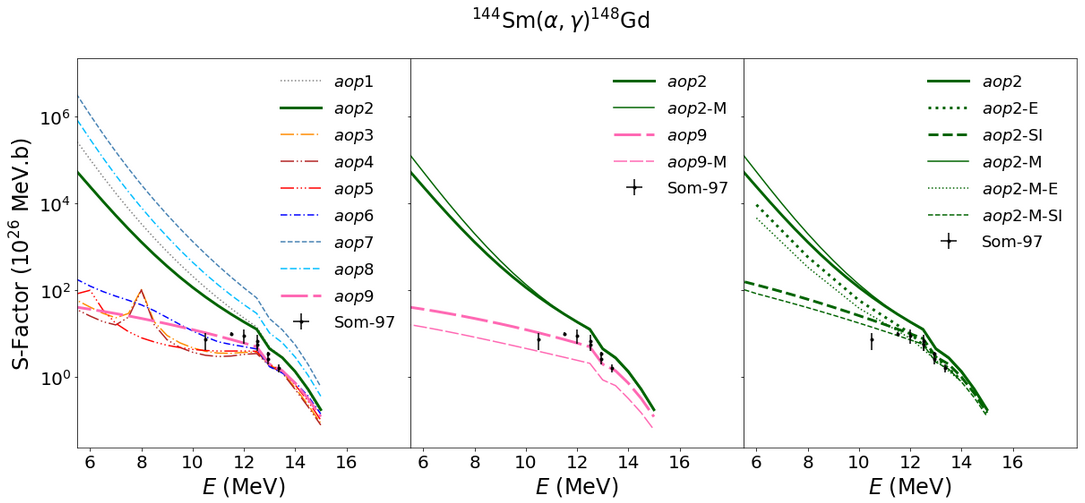}
\caption{
S-factor of $\noyA{Sm}{144}(\alpha,\gamma)\noyA{Gd}{148}$ in the astrophysical energy range ($\sim$ 5 to 12 MeV) extended to the energy range of experimental data from Somorjai-1997~\cite{Somorjai-97}. Left: standard {\AOMP} models. Middle: effect of the fit of {\aomp{2}} and {\aomp{9}} to Mohr-1997 $(\alpha,\alpha)$ data. Right: effect of parameter energy dependence (-E) or sharp imaginary part (-SI); see text for details.
}
\label{fig-SF}
\end{center}
\end{figure}

The $S$-factor for $\noyA{Sm}{144}(\alpha,\gamma)\noyA{Gd}{148}$ is shown in Figure~\ref{fig-SF}, 
together with the data from Somorjai-1997~\cite{Somorjai-97}. 
The $S$-factor $S(E)$ is related to the cross section $\sigma(E)$ by~\cite{Iliadis}:
\begin{equation}
\sigma(E)=\frac{1}{E}S(E)
\exp\left(-2\pi \eta\right)
\end{equation}
where $\eta$ is the Sommerfeld parameter.
The Somorjai-1997 data have shown that traditional {\AOMP} models largely over-estimate the $S$-factor in the astrophysical energy range, as can be seen on the left part of Figure~\ref{fig-SF}. 
Modern {\AOMP} by Demetriou-2002 ({\aomp{3}}, {\aomp{4}}, {\aomp{5}}),
Avrigeanu-2014 ({\aomp{6}}) and Mohr-2020 (Atomki-V2, denoted here {\aomp{9}}) take these constraints into account. 
The middle part of the figure shows that fitting either {\aomp{9}} or the historic {\aomp{2}} on $\noyA{Sm}{144}(\alpha,\alpha)$ data at 20 MeV does not strongly modify their trend, although a sensitive downward shift is observed for {\aomp{9}}. 
On the right part, we focus on modifications made to {\aomp{2}} and {\MSM} for a more accurate description of $(\alpha,\gamma)$ cross sections at low energy.
Originally, the potentials of these two models have no energy dependence.
Two kinds of modifications are considered: 
(i) linear energy dependence of the parameters similar to the one of {\aomp{6}}, denoted here as {\aomp{2}-E};
(ii) constant, sharp imaginary potential similar to the one adopted for {\aomp{9}}, denoted {\aomp{2}-SI}.
For {\aomp{2}-SI}, the $r$ and $a$ parameters of the real part are adjusted to reproduce the same $(\alpha,\alpha)$ angular distribution at 20 MeV as for {\aomp{2}}.
We can see that the sharp imaginary modification (-SI) has a much stronger effect than the parameter energy dependence.
Similar notations are used for the corresponding modifications brought to the {\MSM} model. The corresponding curves for the {\MSM} series are similar to the {\aomp{2}} ones, with slight shifts.

\begin{figure}
\begin{center}
\includegraphics[width=1.0\textwidth]{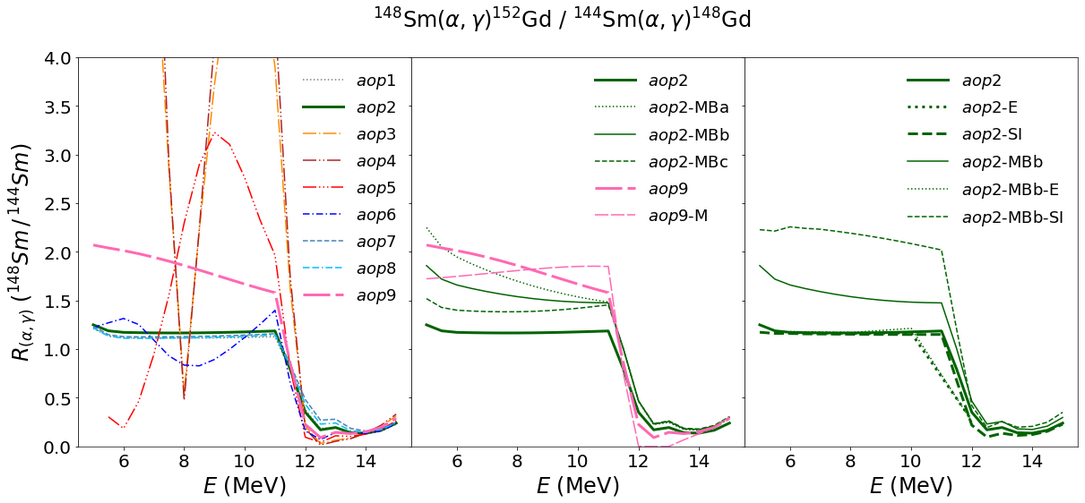}
\caption{Capture cross section isotopic ratio.  
Left: standard {\AOMP} models. 
Some values are out of scale for {\aomp{3}} and {\aomp{4}},
which reach at 10 MeV a local maximum of 5.5 and 6.6 respectively.
Middle: effect of the fit of {\aomp{2}} and {\aomp{9}} to Mohr-1997 $(\alpha,\alpha)$ data, with 3 versions following Badawy-1978 constraints for $\noyA{Sm}{148}$ {\aomp{2}} parameters. Right: effect of parameter energy dependence (-E) or sharp imaginary part (-SI); see text for details.
}
\label{fig-agra}
\end{center}
\end{figure}

Next, we investigate how the {\AOMP} isotopic dependence affects capture cross sections for $\noyA{Sm}{148}$ and $\noyA{Sm}{144}$ by representing in Figure~\ref{fig-agra} the cross section ratio:  
\begin{equation}
\Rag
= \frac
{\sigma (\noyA{Sm}{148}(\alpha,\gamma)\noyA{Gd}{152})}
{\sigma (\noyA{Sm}{144}(\alpha,\gamma)\noyA{Gd}{148})}
\end{equation} 
The sudden reduction around 11 MeV in these plots is due to the opening of the $(\alpha,n)$ channel, which occurs at 10.79 MeV for $\noyA{Sm}{148}$, and 12.25 MeV for $\noyA{Sm}{144}$.
Looking at different standard {\AOMP} on the left part of the figure, we see that most {\WS} models have a common behavior with a plateau around 
$\Rag=1.2$, while {\aomp{6}} presents oscillations around this value.
Larger values of {$\Rag$} are reached with {\DF} models.
The strong variations in the case of Demetriou potentials are linked to resonances that appear for $\noyA{Sm}{144}(\alpha,\gamma)$, 
and not for $\noyA{Sm}{148}(\alpha,\gamma)$: see Figure~\ref{fig-SF}, showing
peaks around 8 MeV for {\aomp{3}} and {\aomp{4}}, and around 6 MeV for {\aomp{5}}.
In the middle part of the figure, we present the effect of {\AOMP} adaptation to the samarium isotopes: fitting $\noyA{Sm}{144}(\alpha,\alpha)$ data at 20 MeV from Mohr-1997, and applying Badawy-1978 constraint given by Eq.~(\ref{eq-Badawy}) for $\noyA{Sm}{148}$ {\aomp{2}} parameters (versions \MSBa, \MSBb, and \MSBc). 
For the {\aomp{2}} series, we observe an increase in $\Rag$ for fitted versions \MSMBa, \MSMBb, and \MSMBc, with an enhancement at low energy for the versions with larger diffuseness of $\noyA{Sm}{148}$ {\AOMP}, namely {\MSBa} and {\MSBb}. 
On the right part of the figure, we consider the effect of parameter energy dependence (-E) and constant sharp imaginary part (-SI), with notations similar to the ones used in Figure~\ref{fig-SF}. 
Let us first consider the original \aomp{2}, whose parameters are identical for $\noyA{Sm}{144}$ and $\noyA{Sm}{148}$.
A modification of this original version by either by -E or -SI procedure has very little effect on the isotopic ratio $\Rag$.
Let us now consider the fitted versions \MSMBb, 
where the potential parameters are different for $\noyA{Sm}{144}$ and $\noyA{Sm}{148}$
and $\Rag$ is increased. 
Making the -E modification, $\Rag$ recovers lower values; 
indeed, the energy dependance described in~\cite{aomp6}
uses a mass-dependent slope for the diffuseness parameter of the imaginary part
such that the difference between $\noyA{Sm}{144}$ and $\noyA{Sm}{148}$ for this parameter is reduced towards lower energy.
In contrast, the -SI modification on {\MSBb} enhances the increase in $\Rag$.

\subsubsection{Impact on \texorpdfstring{$(\alpha,n)$}{(a,n)} reactions below 20 MeV}

\begin{figure}
\begin{center}
\includegraphics[width=1.0\textwidth]{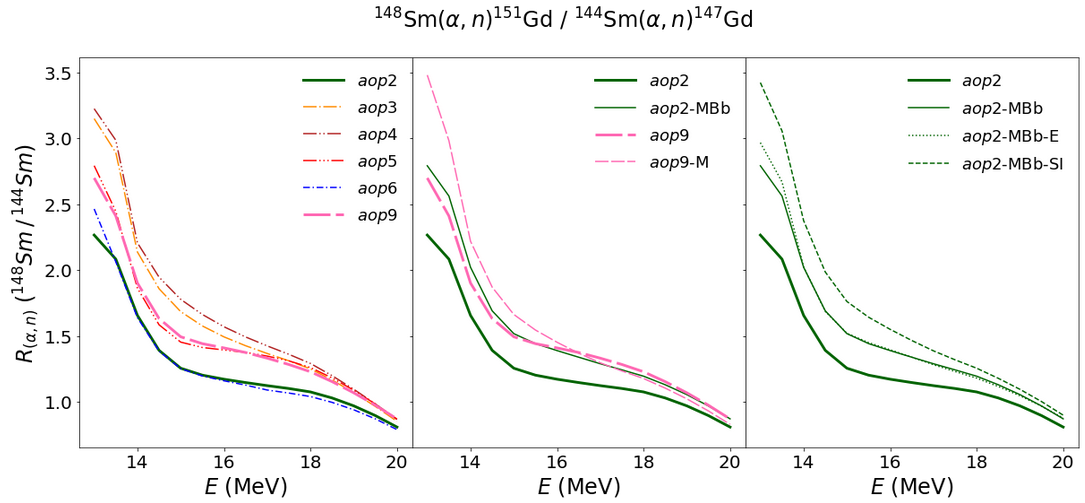}
\caption{Similar to Figure~\ref{fig-agra} in the case of $(\alpha,n)$ reactions below 20~MeV.
}
\label{fig-anra}
\end{center}
\end{figure}

The $\noyA{Sm}{144}(\alpha,n)\noyA{Sm}{147}$ cross section has been recently measured between 13 and 20 MeV by Gyürky \textit{et al.}~\cite{Gyurky-23}. 
As emphasized by the authors, the $(\alpha,n)$ cross-section, as well as the $(\alpha,\gamma)$ cross section below neutron threshold, is essentially sensitive to the $\alpha$ transmission, usually calculated with the optical model. It is then an efficient observable to constrain the {\AOMP}. It seems that no such measurement has been done for $\noyA{Sm}{148}$. 
We show in Figure~\ref{fig-anra} the calculated cross section ratio:
\begin{equation}
\Ran
= \frac
{\sigma (\noyA{Sm}{148}(\alpha,n)\noyA{Gd}{151})}
{\sigma (\noyA{Sm}{144}(\alpha,n)\noyA{Gd}{147})}
\end{equation}
It appears that this ratio would be an important observable to constrain the energy dependence of the {\AOMP} towards the astrophysical energy range. On the left part, we represent this ratio for standard modern {\AOMP}, and the historical one, {\aomp{2}}. 
We see that their results split in three groups: 
the two {\WS} potentials ({\aomp{2}} and {\aomp{6}}) give the lowest ratio, 
while the highest ratio is given by two of the Demetriou potentials ({\aomp{3}} and {\aomp{4}}), 
and intermediate values are obtained for the dispersive Demetriou potential ({\aomp{5}}) and Atomki-V2 potential ({\aomp{9}}). 
In the middle part, we compare standard potentials to their modified versions adapted to samarium isotopes, as introduced previously. 
We see that in both cases ({\aomp{2}} and {\aomp{9}}), the ratio is sensitively increased for the adapted version. 
In the right part, we consider the effect of different treatments applied to the {\AOMP} potential 
with the purpose to describe $\alpha$-induced reactions at low energy:
energy dependence of the parameters (-E) and sharp imaginary part (-SI), as explained previously. We observe that the -E treatment has no impact on the isotopic ratio, while the -SI treatment results in a significant increase.

\section{Conclusion and perspectives}

The study of $\alpha$ elastic scattering isotopic ratios is important to characterize the isotopic dependence of the AOMP.
In the present work, we have measured the scattering of a 20 MeV $\alpha$ beam on two samarium targets of high isotopic purity, $\noyA{Sm}{144}$ and $\noyA{Sm}{148}$, in order to compare the AOMP properties of these nuclei. Angular distributions have been measured between 31$^\circ$ and 85$^\circ$ with the Split-Pole magnetic spectrometer, and between 80$^\circ$ and 163$^\circ$ with silicon detector telescopes. New experimental data have been obtained for relative cross sections: 
the elastic isotopic ratio $\elratio{Sm}{148}{144}$ 
and the inelastic to elastic ratio $\inelratio{Sm}{148}$ (involving inelastic scattering to the first excited state). 

While the case of $\noyA{Sm}{144}$ has already been the focus of many studies,
very few data are available on its closest stable isotope $\noyA{Sm}{148}$.
Since $\noyA{Sm}{144}$ is semi-magic, attention should also be given to non-magic nuclei of this region in order to better constrain the {\AOMP} and obtain a reliable modeling of the nucleosynthetic reaction network in this region, where the $\noyA{Gd}{152}(\gamma,\alpha)\noyA{Sm}{148}$ reaction is in competition with 
$\noyA{Gd}{152}(\gamma,n)\noyA{Gd}{151}$.

Concerning the isotopic ratio for elastic scattering, it is appearant that the {\AOMP} models whose real part is of Woods-Saxon type predict larger values than the ones whose real part is calculated by double-folding. This is reflected by the different behavior of the isotopic ratio of the real part of the potential, especially in the peripheric region. However, it was not possible to extract a specific value of the distance for which the value of this potential ratio would be decisive whatever the potential shape elsewhere. The various characteristics of the {\AOMP} present many correlations, some of which are well-known (such as the so-called continuous ambiguities, see e.g.~\cite{aomp2}), and others are more intricate. 
In this context, machine-learning approaches are promising methods to increase our knowledge of such correlations and their consequences. For instance, a recent study by Marshall et al.~\cite{Marshall-25} uses a bayesian framework to analyse the impact of
$\alpha$ elastic scattering data on $\noyA{Sr}{86}$, of interest for the weak r-process~\cite{Arcones-11}. We also plan to perform bayesian studies in future works, extending the kind of studies we have performed in~\cite{Achment} for nuclear level densities.

Concerning inelastic scattering, the obtained 
values for $\inelratio{Sm}{148}$
are highly precise, with small uncertainties. 
However, such observable does not allow to constrain the {\AOMP}, due to the large theoretical uncertainty on the treatment of inelastic scattering. The coupled-channel approach gives results that differ from the DWBA approach, with strong differences depending on the type of coupling (vibrational or rotational) and the value of the deformation parameter. The best reproduction of our data is obtained for a rotational coupling scheme with the experimental value of the deformation parameter $\beta_2$, although it overestimates the cross section at large angles. For such model, the sensitivity to the {\AOMP} is very weak.
Anyway, a better control of the modeling of inelastic cross section, including a precise value of the deformation parameter, is needed before we can envisage to use it to constrain the {\AOMP}.

We have also considered the impact of the {\AOMP} on $\alpha$-induced reactions at low energy, especially the $(\alpha,\gamma)$ cross section in the astrophysical energy range. 
Both $(\alpha,n)$ and $(\alpha,\gamma)$ below neutron threshold are essentially sensitive to the $\alpha$ transmission probability. 
In both cases, we have observed that the isotopic ratio of the cross sections for 
$\noyA{Sm}{148}$ and $\noyA{Sm}{144}$ increases when we use versions of the {\AOMP} that we have modified to obey constraints from previous experimental data on these nuclei. 
Further, when adding two of the treatments that have proved successful to reproduce the experimental $\noyA{Sm}{144}(\alpha,\gamma)$ cross section at low energy, we have found that they result in different isotopic ratios: 
the parameter energy dependence from~\cite{aomp6} reduces the isotopic ratio for $(\alpha,\gamma)$ and leaves it roughly unchanged for $(\alpha,n)$, while setting a sharp imaginary part like in~\cite{mohr_2020} results in a significant increase of these ratios. 
In the case of $(\alpha,\gamma)$, the two prescriptions result in isotopic ratios differing by roughly a factor of 2 below neutron threshold.
We conclude that it would be very valuable to perform measurements of $\alpha$-induced reactions on $\noyA{Sm}{148}$ at low energy to obtain experimental values of such ratios. This would help to improve our knowledge of the isotopic dependence of the {\AOMP}, address the magicity effect, and better constrain the description of the optical potential toward the astrophysical energy range.


\section*{Acknowledgements}

The authors thank Beyhan Bastin and Stéphane Goriely for fruitful discussions that initiated this project. We thank J. Bourçois, P. Benoît-Lamaîtrie and F. Pallier for the target deposit using the SIDONIE separator of the MOSAIC platform and characterization of the targets. We thank the beam operators of the ALTO facility. 
The authors are grateful to the LABEX Lyon Institute
of Origins (ANR-10-LABX-0066) Lyon for its financial
support within the Plan France 2030 of the French
government operated by the National Research Agency (ANR).
A.~M. S\'anchez-Ben\'{\i}tez thanks economical support from grant EPIT16202023 funded by University of Huelva.



\bibliographystyle{apsrev4-2}
\bibliography{biblio}

\end{document}